\documentstyle[multicol,aps]{revtex}
\begin{document}
\draft
\title{A Scintillating Fiber Hodoscope for a Bremstrahlung
Luminosity Monitor at an Electron$-$Positron Collider}

\author{D.H. Brown, D.H. Orlov, G.S. Varner\footnote{currently at 
Physics Department, University of Hawaii}, W.A. Worstell}
\address{Physics Department, Boston University, 
590 Commonwealth Avenue, Boston, MA 02215}

\author{S.I. Redin\footnote{currently at 
Physics Department, Yale University} }
\address{Budker Institute of Nuclear Physics, Novosibirsk 630090 Russia}

\date{\today}
\maketitle

\begin{abstract}

The performance of a scintillating fiber
(2mm diameter) 
position sensitive detector 
($4.8 \times 4.8$ cm$^2$ active area)
for the single bremstrahlung luminosity monitor 
at the VEPP-2M electron-positron
collider in Novosibirsk, Russia is described. 
Custom electronics is triggered by coincident hits in the 
X and Y planes of 24 fibers each, and reduces 
64 PMT signals 
to a 10 bit (X,Y) address. Hits 
are accumulated (10 kHz) in memory and 
display (few Hz) the VEPP-2M collision vertex.
Fitting the strongly peaked distribution ( $\sim$ 3-4 mm at 1.6m
from the collision vertex of VEPP-2M ) 
to the expected QED angular distribution 
yields a background in agreement with an independent
determination of the VEPP-2M luminosity.

\end{abstract}

\begin{multicols}{2}\narrowtext

\section{Introduction}

The VEPP-2M $e^+ e^-$ collider at the Budker Institute of Nuclear Physics,
Novosibirsk, Russia is currently engaged in high precision measurements of
hadron production in the 1 GeV region. These measurements are useful for
reduction of the uncertainties on hadronic vacuum polarization contributions
(arising from virtual photon interactions with the vector mesons $\rho$,
$\omega$ and $\phi$) to several fundamental constants of physics. Among these
constants are, the muon $g-2$ value \cite{Note220}, the running fine structure
constant evaluated at the $Z$-boson mass, $\alpha(M_Z^2)$ \cite{EJ95}
\cite{S95}, and the hyperfine structure of muonium \cite{STY84}. The detectors
CMD2 \cite{BINP95-35} and SND \cite{BINP87-36} are currently taking data in
VEPP-2M to understand their systematic errors and to make preliminary
measurements for these physics goals and to contribute to the planning of
future high-luminosity $\phi$-factories \cite{BINPhi} \cite{DAFNE}. 

In the CMD2 experiment, precise luminosity determination is performed offline
by analysis of
large angle Bhabha scattering events. Online luminosity determination with
less accuracy is performed by small angle single and double bremstrahlung
monitors.
These are placed outside of CMD2 along tangents to the VEPP-2M colliding beam
ring as shown in Fig.~\ref{f:LM}. The fundamental processes which generate 
photons from the VEPP-2M collision vertex are 
in descending order of cross section:
single and double bremstrahlung, and two photon annihilation \cite{Baier81}.

The CMD2 luminosity monitor (LM) systems consist of $70 \times 70 \times
100$ mm$^3$ BGO crystal scintillators and are
used as rate counters \cite{LMBGO}. The single bremstrahlung signal is
difficult to distinguish from background due to beam-residual gas nuclei
bremstrahlung (which shares the same strongly forward peaked QED angular
distribution as the beam-beam single bremstrahlung signal) and lost beam
interactions in the vacuum chamber. The latter generates photons and charged
particles with a relatively broader angular distribution depending on
VEPP-2M machine conditions. 

The double bremstrahlung signal can be obtained by a method of coincidence 
between signals from LM systems on either side of CMD2. 
The simple coincidence signal
($N_1$) of the two LM systems contains actual double bremstrahlung 
($\gamma \gamma$)
events plus accidental ({\it acc}) coincidences from the
(orders of magnitude larger rate) single bremstrahlung events: 
$ N_1 = N_{\gamma\gamma} + N_{acc} $. 
The accidental coincidence rate can be measured by the coincidence
signal ($N_2 = N_{acc}$) between one LM signal delayed by a few VEPP-2M beam
crossings from the undelayed signal in which double bremstrahlung events
cannot possibly occur.
The accidental background can then be
removed by subtraction of the two which isolates 
the double bremstrahlung events: 
\begin{eqnarray*}
N_{\gamma\gamma} &=& N_1 - N_2 \\
       &=& (N_{\gamma\gamma} + N_{acc}) - N_{acc} 
\end{eqnarray*}
The problem with this method is that the small 
double bremstrahlung cross section implies a small 
number of events which means a large
statistical error.
This leads in practise to occasional negative
total counts for double bremstrahlung. 

In response to the failure of rate-dependent techniques to distinguish
between luminosity monitoring processes and backgrounds, the possibility to
identify a signal process by its characteristic angular distribution can then
be exploited. For this purpose a small diameter scintillating fiber hodoscope
is employed as a
luminosity monitor position sensitive detector
(LMPSD) for the incidence face of the LMBGO to measure the sharply peaked
profile of the photon distribution. The deviation from the expected QED angular
distribution is used to measure the background to be subtracted
from the total LMBGO signal. In addition, the LMPSD information 
provides real time visual feedback for accelerator control of the position and
stability over time of the VEPP-2M colliding beam vertex 
(see Fig.~\ref{f:TVview}).

The organization of this note is as follows.
Sec.~\ref{s:components} outlines the main LMPSD components.
Sec.~\ref{s:performance} describes the uniformity measurements taken with a
radioactive source and the performance of the LMPSD during colliding-beams
operation of VEPP-2M. 

\section{LMPSD Components} \label{s:components} 

The LMPSD consists of a scintillating fiber package mounted to a PMT-base
assembly with custom designed readout electronics. The fiber package, PMT-base
assembly and readout electronics are discussed below. 

\subsection{Scintillating Fiber Package} 

Two scintillating fiber planes each have 24 fibers and are oriented in
transverse directions to the incident photons from the
interaction region of VEPP-2M. The active part of the fiber planes are 2mm
diameter
round radiation-hard Bicron scintillator each 6cm in length. They are
index-of-refraction matched by Bicron BC-600 epoxy to 25cm long 2mm round
non-scintillating cladded polystyrene fibers. An aluminum frame has been
constructed to fix the fibers without introducing
material before or after the fiber planes in the incidence window. 

Both ends of the fibers were polished. 
The fiber ends 
in the aluminum frame are held in place by clamps and adhesive to a mirrored
surface, which enhances by 60 \% the un-mirrored fiber light yield. 
The PMT-end
of the fibers are held in place by a grid of 64 guide holes machined in a
lucite plate. 
Cross
talk was minimized by moving the plate within the tolerance of the bolt holes
and measuring signals in a 3x3 grid of pixels with a UV-excited test fiber in
the central pixel. 
This guide-plate is pinned to the PMT mounting plate.
The entire fiber/PMT package is shown schematically in Fig.~\ref{f:hardware}.

\subsection{PMT Base Assembly} 

The photomultiplier (PMT) used is a Philips XP4722 segmented output
electrode device with 64
individual 10 stage electron-multiplication channels. 
The PMT base assembly was custom
designed and built in a circular array which allows each channel to be
amplified separately by a current feedback operational amplifier
PMI OP-160. All channels were mounted on a radial spoke (32 per side) of the
circular printed circuit board.
% shown in Fig.~\ref{f:pcboard}. 
The channels on one side were rotated relative to the channels on the other to
minimize cross talk through the pc board. 

After initial
testing the PMT gains 
were equalized by modifiying the op-amp feedback resistors.
Two raised common voltage rings supply the op-amps. 

\subsection{Mechanical Support} 

The fiber package is supported by attachments to the PMT mounting plate.
This plate is
insulating since the Philips XP4722 is used with the exterior metal
ring at high voltage to maximize the PMT gain. A clear lucite fiber guide plate
(with an 8x8 grid of fiber feedthrough holes) is fixed to the black plastic
PMT mounting plate. The non-scintillating part of the fibers are bent equally
from the two perpendicular planes in the fiber package to the PMT. 

The other side of the PMT plate has a groove which accomodates the rim of a
tapering (5" - 2") mu-metal shield which provides a light tight enclosure for
the PMT base assembly.
The cables are run out the narrow end of the mu-metal shield which is suitably
taped closed for light tightness. 

The amplified signal from each channel is conducted 2 meters through
individual RG-174 coaxial cable to the ``CODER'' readout electronics.
The CODER is located in  
a special NIM crate mounted under the CMD2 
detector which converts the 64
signals to a 10 bit address for the LMPSD hit (5 bits for X, 5 bits for Y). The
10 bit address is then sent by a 4 meter 40 pin ribbon cable to a CAMAC crate. 

\subsection{The CODER Readout Electronics} 

The CODER electronics 
consists of commercial discriminators, three custom 
NIM modules and a custom CAMAC unit
and is shown in Fig.~\ref{f:CODER}.
The first NIM module 
delivers two 5-bit
values, by means of 32-bit priority encoders, for the HI and LO
fibers excited in a given plane. A second such module produces HI-LO values
for the other fiber plane.
If HI$-$LO$=$0 then only a
single fiber was hit. If HI$-$LO$>$0 then there were multiple fibers hit in the
plane. (In VEPP-2M, the rate for multiple hits turned out to be quite small.)
The HI-LO modules also issue
pulses useful for determining
the timing of the hit. 

The third NIM module is the CODER proper which delivers the HI$-$LO$=$DIFF
difference and the (HI$+$LO)/2$=$MEAN center of the hit by means of a 2k 
PROM. A dip switch defining
N allows the selection of events with DIFF$<$N. The signals are transformed
to differential TTL to ensure that signal degradation to the CAMAC crate
will not destroy LMPSD information. The CAMAC unit transforms the signals
back to ground 
relative TTL and defines the valid event condition described below.

The CODER data output consists of two 16-bit words, one for X and the other
for the Y plane. Each word is issued in parallel upon a valid event
defined by the logical AND between three signals: X plane OR, Y plane OR,
and computer-not-busy. This signal defines the CAMAC LAM.
The least significant four bits (of the 16-bit data word for each fiber plane)
are the DIFF, 
the next more significant five bits are the MEAN
taken as the photon position, while the most significant two label the
data word as X or Y. The computer program sends the DIFF and MEAN bits (along
with the VEPP-2M beam current updated every 100th event) 
to a disk resident file in a form suitable for analysis software \cite{PAW}.
The logical AND signal of the X and Y plane ORs (without
computer-busy veto) was additionally readout and sent to a CAMAC scaler. 

\section{ LMPSD Performance }
\label{s:performance}

\subsection{Calibration}

Before mounting the LMPSD in VEPP-2M, 
LMPSD signal responses to illumination by a \hbox{$Ru^{106}$}
source were measured to determine (and equalize) the uniformity among
fiber channel responses throughout the data acquisition chain.

The 64-channel PMT-base assembly has gain variations of at least a
factor of 2 while there are also 25\% variations in individual
fiber light yields. The discriminator thresholds in the CODER 
of the fiber-PMT
signals provide a convenient means for equalizing the
channel responses to constant illumination. Constant illumination is 
achieved with an uncollimated \hbox{$Ru^{106}$} source of 3.5 MeV electrons
placed 15 cm away from the X and Y planes of the LMPSD.

There are 4 discriminator thresholds for the 48 channels (24 for 
both X and Y) in the LMPSD. After checking the signal response 
to the collimated \hbox{$Ru^{106}$} source 
for each channel on an oscilloscope the channels were grouped
into 4 threshold groups. The LMPSD was subjected
to uncollimated 
constant illumination and the thresholds were adjusted to improve
the uniformity of response across the channels. In some cases,
high gain PMT channels were re-matched with low light yielding fibers
to create a more uniform response.
%
%Figure 1 shows a signal response with 25\% variations for the X plane
%in two plots a week apart to show reproducability of the measurements.
%The upper plot in Figure 2 shows the Y plane uniformity to be $>$ 50\%
%with a strong Y1 and weak Y16 channel, and in the lower plot 30 \%
%uniformity is achieved after switching the position of fibers Y1 and
%Y16 on the PMT pixels. Figure 3 shows the 30 \% uniformity over a month
%later after the beam-residual gas single-bremstrahlung tests and a further
%reinstallation on the electron side of CMD2. 
The uniformity %therefore
appears to be quite stable over long periods of operation and after complete
reinstallation procedures (from one side of CMD2 to the other).

The channels tuned in this manner
have yielded the photon beam position in the data
presented below and appear to be acceptable. 
The limitation to the uniformity is determined mainly
by the lack of individual channel discrimination. If this were built
into the pc-board or if more discriminators were available the uniformity
of the device could be improved. 

\subsection{Conversion of Photons}

As depicted by small wedges on the outer radii of the bending magnets 
in Fig.~\ref{f:LM}, the vacuum chamber of the VEPP-2M
storage ring has special photon channels with a thin ($\sim 1$ mm)
stainless steel
window ($\sim 2 \times 4$ cm$^2$) through which the
bremstrahlung photons may pass. In principle, single bremstrahlung
photons will convert to an 
$e^+e^-$ pair through this chamber wall.
However, earlier measurements using wire chambers determined that beam
loss interactions in the vacuum chamber and other hardware
in the VEPP-2M ring cause a much larger
charged particle background at the luminosity monitors
than the photon-$e^+ e^-$ conversion signal.

Therefore, a scintillator/PMT assembly was placed just in front of
the LMPSD to veto charged particles. This was implemented in the electronics
by taking both of the discriminator outputs from the scintillator
signal and feeding them into channel 30 and 32 in the HI-LO
NIM modules. With DIFF$<$1 event definition, this meant that a
scintillator hit in those channels {\it and} a fiber hit in 
channels 1-24 would certainly yield DIFF$>$1 and hence the event would
not be issued as valid.

With photon conversions in the vacuum chamber window
and {\it any} other charged particle background mostly removed 
by a scintillator veto, photon purity of the signal
reaching the LMPSD was significantly increased.
To introduce post-scintillator conversion and thereby increase
the efficiency of the LMPSD several different thicknesses
of tungsten plates 
(varying from 1 mm to 3.5 cm) 
were tested. In addition, the light tight enclosure built around the
fiber package introduces a further 1 mm of stainless steel.
After comparison of the
peak value and width of the two dimensional distributions 
for all configurations
(the largest peak results in the
narrowest width because of the self-trigger of LMPSD)
the optimal configuration was observed to be an additional 1 mm stainless steel
plate placed in front of the LMPSD enclosure.

\subsection{Colliding Beam Data}

The LMPSD is intended to provide position information useful 
for determining the angular distribution of the single bremstrahlung signal.
In the operation of the LMPSD, the photons from 
the VEPP-2M collision vertex are incident 
upon the two scintillating fiber planes located (after an 
$e^+ e^-$ conversion plate, see above)
in front of a BGO crystal electromagnetic calorimeter. 
The luminosity determination by the single bremstrahlung process
can be improved by removing 
the deviation from the expected QED angular distribution
which falls off faster than the LMPSD data.

The LMPSD signals were accumulated for 10k events and read out
to a CAMAC memory register (CMR) by PDP-11 type and 
DEC VAX assembler. The signals from the CMR, were sent both to computer memory
and through a Color Display Unit to an online color monitor
for control room viewing, as in Fig.\ref{f:TVview}. 
A clean collision vertex is characterized by a single bremstrahlung peak;
in comparison, noticeable changes (additional peaks) were observed
with deliberate VEPP-2M orbit modifications (transverse 
displacement by a few mm).

The LMPSD signal for VEPP-2M operation with colliding beams
at energy 497 MeV per beam
is shown in 3D relief for several configurations in
Figs.~\ref{f:LMPSD-alone}-\ref{f:LMPSD-bkgd}. 
Fig.~\ref{f:LMPSD-alone} is the LMPSD signal with self $XY$-trigger alone;
Fig.~\ref{f:LMPSD-veto} is the LMPSD
signal in anti-coincidence with a scintillator paddle
in front of the LMPSD to veto charged particle incidence;
Fig.~\ref{f:LMPSD-BGO} is the LMPSD/scintillator-veto signal in further
coincidence with the BGO calorimeter signal above
an energy threshold of 200 MeV; all have the same total number of events
and for comparison have the same vertical scale.
Clearly, the scintillator-veto removes charged particle
background and the BGO coincidence
removes remaining low energy photon background; both combinations
increase the LMPSD single bremstrahlung signal relative to background.
These distributions confirm earlier measurements with a multiwire
proportional chamber
which indicated that the beam loss background was 
predominantly low energy in composition and broadly distributed 
\cite{LMBGO}. 

Without coincidence, the self-trigger
of the LMPSD signal alone
is not sufficient to observe the QED angular distribution
of the single bremstrahlung signal; the self-trigger signal distribution
is almost as populated off the bremstrahlung peak
due to the low energy background as it is on peak. 
Therefore, the best operating
conditions of the LMPSD is in conjunction with both the
scintillator veto and BGO signal above threshold. This is the data
which is fit by the expected QED angular distribution.

\end{multicols}\widetext

\subsection{Fit to Quantum Electrodynamics}

The final LMPSD/Veto/BGO signal can now be compared with the
QED angular distribution of the single bremstrahlung process:
$$
\sigma_{\gamma} (x,y) =  { {\rm d}^2 \sigma_{\gamma} \over {\rm d}x {\rm d}y }
= {\gamma^2 \over 2 \pi} 
\left[ 
E_1 { 1 \over 
   \left( 1 + {\gamma^2 \over z^2 } \left( x^2 + y^2 \right) \right)^2 } +
E_2 {         {\gamma^2 \over z^2 } \left( x^2 + y^2 \right) 
        \over 
   \left( 1 + {\gamma^2 \over z^2 } \left( x^2 + y^2 \right) \right)^4 } 
\right] 
$$

\begin{multicols}{2}\narrowtext

\noindent
where $\gamma$ is the electron Lorentz factor, $z$ is the distance
from the VEPP-2M collision vertex to the LMPSD, and
$x,y$ are the transverse dimensions of the LMPSD hodoscope window.
The explicit form of the
dimensionless energy and threshold dependent constants $E_1$ and $E_2$ 
are given in Appendix~\ref{a:QED}.

An online result from the LMPSD requires
simplification of the true QED formula 
in order to minimize dead time for the CMD2 experiment. A simplified form 
which exhibits the essential behaviour of
the principal first term of the QED angular distribution is:
$$
\sigma_{\gamma} (x-x_0,y-y_0) \simeq { A \over 
   \left( 1 + a \left( x - x_0 \right)^2 + 
              b \left( y - y_0 \right)^2    \right)^2 } .
$$
The parameters $a$ and $b$ characterize the width of
the distributions in the two dimensions and are allowed to vary in the fit.
The five parameters
of the fit are: amplitude $A$, center $(x_0,y_0)$ and widths $(a,b)$
of the distribution.

The procedure for determination of the background is to fit
the data
to the expected QED distribution
only at 
the top part of the peak 
where the signal is expected to
dominate. Next, the optimized fit parameters are used to
extrapolate the QED function to the tail regions
where the background is expected to be larger.
Such a fit (3 fibers in the $x$ and $y$ planes each)
is shown in Fig.~\ref{f:LMPSD-QED} for the data
of Fig.~\ref{f:LMPSD-BGO}; 
a one dimensional slice of the two dimensional distribution is shown in
Fig.~\ref{f:LMPSD-sliceON} on peak
and in Fig.~\ref{f:LMPSD-sliceOFF} off peak;
while the fit on peak is quite good for a range of
fibers included in the fit (depicted by the several curves shown), 
the background above the QED
fit is clearly seen in the off peak slice.

By plotting the QED function with the fit parameters from the
peak over the whole area of LMPSD the number of total
events due to the QED single bremstrahlung process can be determined.
The remaining events 
are considered background. A plot of the background
which is the difference of the data and the full QED plot is given
in Fig.~\ref{f:LMPSD-bkgd}. The relative number of events in 
background and QED signal 
for four different fit regions are presented below
in Tab.~\ref{t:5fit} for a five parameter fit.
The fit regions are
defined by how many fibers 
relative to the maximum pixel $(x_0,y_0)$ 
in the LMPSD window 
are included in the fit.

\subsection{Comparison of LMPSD with LMBGO and CMD2}

The LMPSD results for the background in the LMBGO monitors may now be compared
with results derived from independent luminosity measurements by CMD2 and
LMBGO. The CMD2 luminosity measurement for Run 2729 is determined by the number
of and QED cross section for Bhabha scattering events to be 
$ {\cal{L}}_{CMD2} = N_{ee} / \sigma_{ee} =$ 6.2 $\pm$ 0.12 ${nb}^{-1}$. 
The quoted error is statistical only.

The LMBGO determination of the VEPP-2M luminosity is based on
a similar relation using the QED cross section for single bremstrahlung events: 
$ {\cal{L}}_{LMBGO} = N_{LMBGO} / \sigma_{\gamma} $
which for CMD2 Run 2729 is
calculated to be $\sim$ 7.7 ${nb}^{-1}$.
The statistical error is negligible compared to the systematic
error due to presence of background which is the object under study.
Comparison with the CMD2 luminosity yields
a first approximation to the amount of background in the LMBGO
luminosity as follows: 
$$
{ {\cal{L}}_{LMBGO} \over {\cal{L}}_{CMD2} } =
{ {\cal{L}}_{\gamma} + {\cal{L}}_{bkgd} \over {\cal{L}}_{CMD2} } =
1 + { {\cal{L}}_{bkgd} \over {\cal{L}}_{\gamma} } = 1.24 .
$$
It is assumed that the ideal single bremstrahlung and CMD2 Bhabha
luminosity are equal: ${\cal{L}}_{\gamma} = {\cal{L}}_{CMD2} $.

The ratio of background to single bremstrahlung signal obtained
in this way should be equivalent to the ratio of background and 
single bremstrahlung signal obtained by LMPSD: 
$$
{ N_{LMPSD} \over N_{\gamma} } =
{ N_{\gamma} + N_{bkgd} \over N_{\gamma} } =
1 + { N_{bkgd} \over N_{\gamma} } = 1.25 \pm 0.05
$$
where the ratio from Tab.~\ref{t:5fit} Fit 3 is shown for comparison.
The systematic error is determined by variation of the QED fit region
on the peak of the LMPSD data. As seen in Tab.~\ref{t:5fit}, the
ratio of background to signal events varies from 20\% to 28\%
for the four different fits presented there. 

The implication of this result is that the LMPSD can produce
a result ON-line for how much of the LMBGO signal needs to be
subtracted for a more accurate luminosity measurement.
The luminosity measurements from the LMBGO and CMD2 above
were determined long after data taking by OFF-line processing.
The LMPSD can provide its measurement of the background to signal
ratio in each monitor record written to tape {\it during}
data taking. In addition, the LMPSD-corrected LMBGO result
can be simultaneously displayed ON-line. This information
and the automatically refreshing visual beam spot display from the 
LMPSD system
are useful
for shift operators to determine both the quality of beam collisions and to
control more precisely how much luminosity
is taken by CMD2 at each energy point of the data taking scan.

%The corresponding LMPSD ratios are presented
%in Tab.~\ref{t:5fit}. Fit 1 and Fits 2 - 5
%all yield a single bremstrahlung to background too large
%since the $3\times 3$ and $3\times 4$ fiber ranges are too small
%or include too much on the large background side
%of the single bremstrahlung peak. Notice in Fig.~\ref{f:LMPSD-BGO}
%that there is a significant amount of background particularly on the
%lower $Y$-fiber side of the peak. This means that Fit 4 should
%include too much background and not be a good fit.
%Observe the opposite is the case as shown in Tab.~\ref{t:5fit}. 

\bigskip\noindent
{\it Acknowledgements}
\medskip

The authors would like to thank Drs. Boris I. Khazin,
Andrei G. Shamov, Vladimir P. Smaktin and Dmitri N. Grigoriev
of the Budker Institute of Nuclear Physics
for proposing the need for the LMPSD, providing support at VEPP-2M
for data aquisition and analysis of results.

\appendix

\section{QED Angular Distribution for Single Bremstrahlung}
\label{a:QED}

There are eight Feynman diagrams which contribute to the
single bremstrahlung process in Bhabha scattering: four 
(one radiated photon for each external particle) associated
with the $s$ channel, and another four with the $t$ channel process.
For application to photons being detected with the CMD2 luminosity
monitor system, the general calculation 
\cite{Baier81} 
is simplified for the limit of photons emitted
close to the incident beam axis and at ultrarelativistic electron energies.
In this limit, only the four $t$ channel diagrams contribute
and the double differential cross section with respect to
polar angle $n=\theta \gamma=\theta E / m_e$ and 
photon energy $\omega$ (energy fraction $x \equiv \omega/E$) is given by
\cite{R90}:
\begin{eqnarray*}
{ {\rm d}^2 \sigma_{\gamma} \over {\rm d} \omega  {\rm d} n }
&=& 4 \alpha \ r_0^2 \ {1 \over \omega} \ { n \over ( 1 + n^2 )^2} \ 
\left[ 
I_1 + ( I_2 + x^2 ) \ln \left( 4 \gamma^2 \left\{ 1/x - 1 \right\} \right) 
\right] \\
I_1 &=& - ( 2 - x )^2   + { 16 n^2 \over ( 1 + n^2 )^2 } ( 1 - x )  \\
I_2 &=& \ 2 ( 1 - x ) \ - {  4 n^2 \over ( 1 + n^2 )^2 } ( 1 - x )   
\end{eqnarray*}
where $r_0$ is the classical electron radius.

Since LMPSD will ultimately be used in conjunction with
the BGO calorimeter system which 
detects photons above an energy threshold $\omega_{th}$
(to remove low energy backgrounds),
this expression needs to be integrated from $\omega_{th}$ to
$E \equiv E_{max}^{\gamma} \simeq E_{beam}^e$ in order to obtain the
QED angular distribution of the single bremstrahlung photons.
It is convenient to write $\omega= \omega_{th}$,
$x = x_{th}$ for notation in the result below:

\end{multicols}\widetext

\begin{eqnarray*}
\int_{\omega}^E {\rm d} \omega
{ {\rm d}^2 \sigma_{\gamma} \over {\rm d} \omega {\rm d} n} =
{ {\rm d} \sigma_{\gamma} \over {\rm d} n } &=& 
4 \alpha \ r_0^2 \left[ { n \over ( 1 + n^2 )^2} \ 
\left( 4 A + 2 B + C + Z_2 \right) +
{ n^3 \over ( 1 + n^2 )^4} \ 
\left( 16 A - 4 B \right) \right] \\
Z_n &=& {1 \over n^2} \left( x^n - 1 \right) \\
A &=& \ln(E-\omega) + Z_1 \\
B &=& \ln{4 \gamma^2} \ln(E-\omega) + \ln{E} \ln(E+\omega) 
{1\over 2} \left( (\ln E)^2 - (\ln \omega)^2 \right) + \\
& & \ 
x \ln \omega - \ln E - Z_1 \ln { E - \omega \over 4 \gamma^2 } +
\sum_1^{\infty} Z_n \\
C &=& Z_2 \ln{\omega \over 4 \gamma^2 \left( E - \omega \right)}
+ {1\over 2} \left( \ln{\omega} - \ln{E} \right) - {1\over 2}
\end{eqnarray*}

\begin{multicols}{2}\narrowtext

In order to write this in terms of the LMPSD hodoscope variables
$x,y$ it is necessary to make a small angle approximation
since $x,y < 5$ cm and $z=1.6$ m 
(the distance from the CMD2 collision vertex to LMPSD)
and note that:
$$
n = \gamma \theta \to n {\rm d}n = \gamma^2 \theta 
{\rm d}\theta { {\rm d}\phi \over 2 \pi }
\simeq { \gamma^2 \over 2 \pi } \sin \theta {\rm d}\theta {\rm d}\phi
     = { \gamma^2 \over 2 \pi } {\rm d}x {\rm d}y
$$
$$
\theta \simeq \sin \theta \simeq {r \over z} = 
{ \sqrt{ x^2 + y^2 } \over z } \to n^2 = \gamma^2 \theta^2 
\simeq {\gamma^2 \over z^2 } \left( x^2 + y^2 \right) .
$$
Making these substitutions the result is as quoted:

\end{multicols}\widetext

\begin{eqnarray*}
\sigma_{\gamma} (x,y) =  { {\rm d}^2 \sigma_{\gamma} \over {\rm d}x {\rm d}y }
&=& {\gamma^2 \over 2 \pi} 
\left[ 
E_1 { 1 \over 
   \left( 1 + {\gamma^2 \over z^2 } \left( x^2 + y^2 \right) \right)^2 } +
E_2 {         {\gamma^2 \over z^2 } \left( x^2 + y^2 \right) 
        \over 
   \left( 1 + {\gamma^2 \over z^2 } \left( x^2 + y^2 \right) \right)^4 } 
\right] \\[2mm]
E_1 &=& 4 A + 2 B + C + Z_2  \\
E_2 &=& 16 A - 4 B  .
\end{eqnarray*}

\begin{multicols}{2}\narrowtext

%For beam energy $E=497$ MeV and various thresholds 
%these contants are listed in Tab.~\ref{t:e1e2}.

%  \bibliography{lmpsd_ref}
%  \bibliographystyle{unsrt}

\end{multicols}\widetext

\newpage

%\begin{table}
%\caption{The beam energy and threshold depedendent constants in the
%QED angular distribution for single bremstrahlung events.}
%\label{t:e1e2}
%\begin{tabular}{|c|c|c|}
%$\omega$ [MeV] & $E_1$ & $E_2$ \\
%\end{tabular}
%\end{table}

\begin{table}
\caption{Results of a five parameter fit. Total number of events
is 9889. Useful result is ratio of background to signal events.}
\label{t:5fit}
\begin{tabular}{|c|c|c|c|c|c|c|c|c|}
 Fit  
 & $X$-region 
               & $Y$-region
                             & $N_{pixels}$ 
                                & $\chi^2$ 
                                        & $N_{\gamma}$ 
                                                 & $N_{bkgd}$ 
                                                        & $N_{bkgd}/N_{\gamma}$ 
                                                                 & $C$ \\ \hline
%1& $x_0-1 : x_0+1$ & $y_0-1 : y_0+1$ &9 &7.4240 &6340.9 &3548.1 &0.5596 & - \\ 
%2& $x_0-2 : x_0+1$ & $y_0-1 : y_0+1$ &12&5.1879 &6798.5 &3090.5 &0.4546 & - \\
%3& $x_0-1 : x_0+2$ & $y_0-1 : y_0+1$ &12&6.2627 &7586.8 &2302.2 &0.3034 & - \\
%4& $x_0-1 : x_0+1$ & $y_0-2 : y_0+1$ &12&6.3449 &7563.6 &2325.4 &0.3074 & - \\
%5& $x_0-1 : x_0+1$ & $y_0-1 : y_0+2$ &12&6.8239 &6955.0 &2934.0 &0.4218 & - \\
1& $x_0-2 : x_0+2$ & $y_0-1 : y_0+1$ &15&4.7567 &7744.9 &2144.1 &0.2768 & - \\
2& $x_0-1 : x_0+2$ & $y_0-2 : y_0+1$ &16&4.4279 &8075.4 &1813.6 &0.2246 & - \\
3& $x_0-1 : x_0+1$ & $y_0-2 : y_0+2$ &15&5.8754 &7902.8 &1986.2 &0.2513 & - \\
4& $x_0-2 : x_0+2$ & $y_0-2 : y_0+2$ &25&3.3544 &8238.5 &1650.5 &0.2003 & -    
\end{tabular}
\end{table}

%\begin{table}
%\caption{Results of a six parameter fit. Total number of events
%is 9889. Useful result is ratio of background to signal events.
%Fits 6 and 8 had minor convergence problems in MINUIT (parameter errors at
%limits and error matrix not positive definite).}
%\label{t:6fit}
%\begin{tabular}{|c|c|c|c|c|c|c|c|c|}
% Fit  
% & $X$-region 
%               & $Y$-region
%                             & $N_{pixels}$ 
%                                & $\chi^2$ 
%                                        & $N_{\gamma}$ 
%                                                 & $N_{bkgd}$ 
%                                                       & $N_{bkgd}/N_{\gamma}$ 
%                                                                 & $C$ \\ \hline
%1& $x_0-1 : x_0+1$ & $y_0-1 : y_0+1$ &9  &9.8987 &6340.6  &3548.4 &0.5596   &0.000  \\
%2& $x_0-2 : x_0+1$ & $y_0-1 : y_0+1$ &12 &5.7856 &5755.9  &4133.1 &0.7181   &20.563 \\
%3& $x_0-1 : x_0+2$ & $y_0-1 : y_0+1$ &12 &7.3065 &7587.6  &2301.4 &0.3033   &0.000  \\
%4& $x_0-1 : x_0+1$ & $y_0-2 : y_0+1$ &12 &7.4024 &7563.1  &2325.9 &0.3075   &0.000  \\
%5& $x_0-1 : x_0+1$ & $y_0-1 : y_0+2$ &12 &7.5927 &5571.7  &4317.3 &0.7749   &27.704 \\
%%6& $x_0-2 : x_0+2$ & $y_0-1 : y_0+1$ &15 &5.2827 &7604.2  &2284.8 &0.3005   &2.407  \\
%7& $x_0-1 : x_0+2$ & $y_0-2 : y_0+1$ &16 &4.8707 &8073.9  &1815.1 &0.2248   &0.000  \\
%%8& $x_0-1 : x_0+1$ & $y_0-2 : y_0+2$ &15 &6.5282 &7902.3  &1986.7 &0.2514   &0.000  \\
%9& $x_0-2 : x_0+2$ & $y_0-2 : y_0+2$ &25 &3.5189 &7934.5  &1954.5 &0.2463   &4.501      
%\end{tabular}
%\end{table}

\begin{figure}
\caption{Schematic of the Bremstrahlung Luminosity Monitor Systems
in VEPP-2M. Two BGO crystals $70 \times 70 \times 100 mm^3$
are indicated as LMBGO, while the two planes of scintillating
fibers in front of LMBGO are indicated as LMPSD.}
\label{f:LM}
\end{figure}

\begin{figure}
\caption{Schematic of Luminosity Monitor Position Sensitive
Detector (LMPSD) hardware. Scintillating fibers are spliced to
clear fibers (by index-matching epoxy) and mounted on 64 channel
PMT in light tight insulating enclosure.}
\label{f:hardware}
\end{figure}

\begin{figure}
\caption{Schematic for the LMPSD ``CODER'' electronics.
HI-LO determination performed by four 32 bit priority encoders,
while mean and differences are burned into two 2K PROMs.
Events defined by zero difference are read out by CAMAC.}
\label{f:CODER}
\end{figure}

\begin{figure}
\caption{View of LMPSD in the CMD2 Control Room shows two
dimensional colour of single beam spot. VEPP-2M orbit changes
visibly widen and/or produce additional beam spots.}
\label{f:TVview}
\end{figure}

\begin{figure}
\caption{The LMPSD data with self $XY$-trigger alone.}
\label{f:LMPSD-alone}
\end{figure}

\begin{figure}
\caption{The LMPSD data in anti-coincidence with an upstream scintillator.}
\label{f:LMPSD-veto}
\end{figure}

\begin{figure}
\caption{The LMPSD data in anti-coincidence with an upstream scintillator
and in coincidence with the BGO calorimeter above 200 MeV threshold.}
\label{f:LMPSD-BGO}
\end{figure}

\begin{figure}
\caption{The best fit of the LMPSD data with the QED angular distribution.}
\label{f:LMPSD-QED}
\end{figure}

\begin{figure}
\caption{A slice of the LMPSD data with fit: on peak.
The many curves are for the fit regions shown in Tab.~\ref{t:5fit}
and show agreement with the data.}
\label{f:LMPSD-sliceON}
\end{figure}

\begin{figure}
\caption{A slice of the LMPSD data with fit: off peak.
The many curves are for the fit regions shown in Tab.~\ref{t:5fit}
and show how much data is due to signal (under) and background (above curves).} 
\label{f:LMPSD-sliceOFF}
\end{figure}

\begin{figure}
\caption{The background in the LMPSD data: difference of data with fit.}
\label{f:LMPSD-bkgd}
\end{figure}

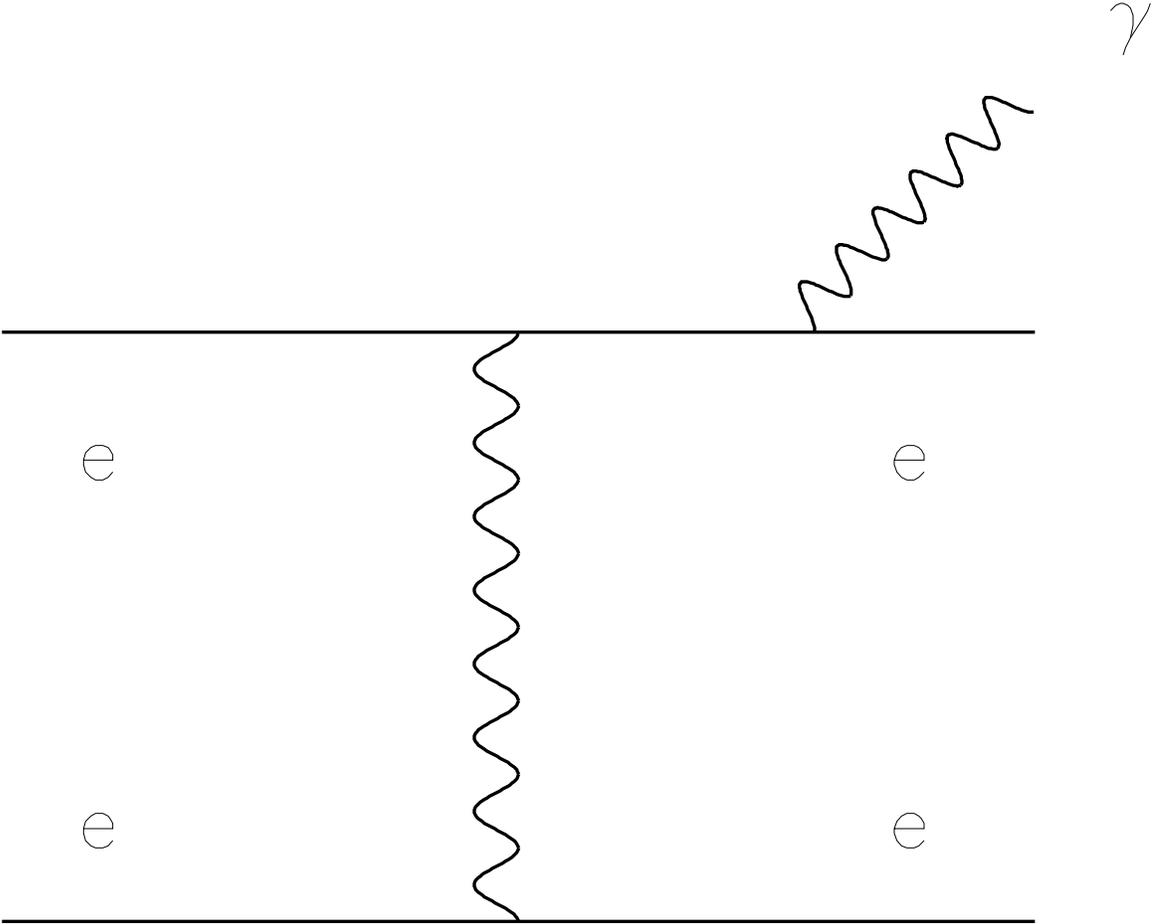
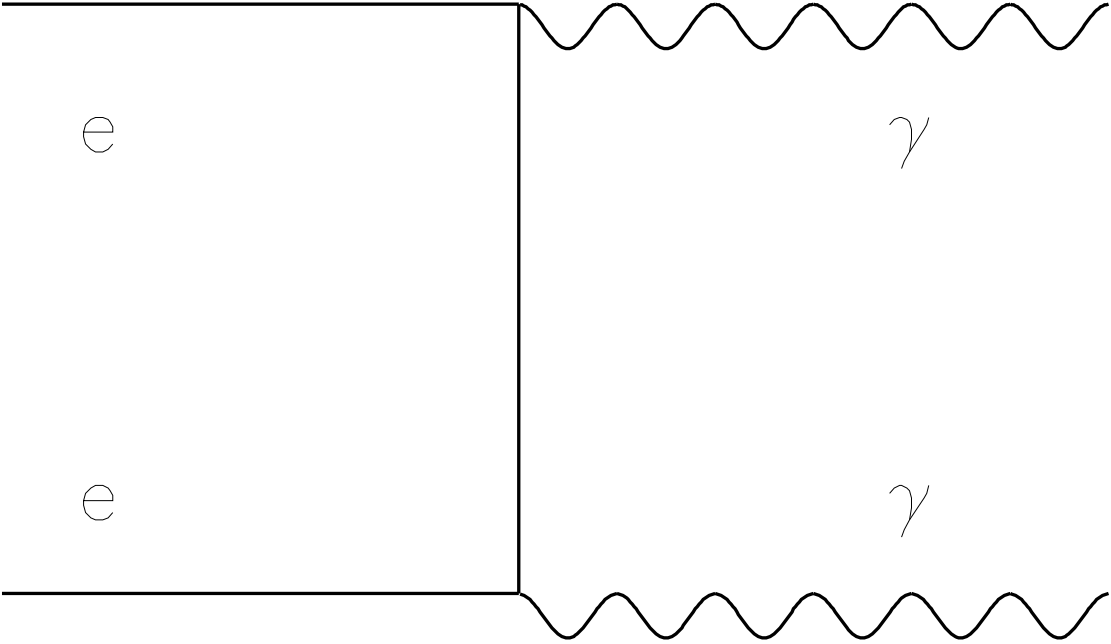
\end{document}